\documentclass[twocolumn,prb,showpacs,superscriptaddress,preprintnumbers,amsmath,amssymb]{revtex4}

\usepackage[dvips]{graphicx}
\usepackage{dcolumn}
\usepackage{bm}

\begin{document}


\title{Crystal-fields in YbInNi$_4$ determined with magnetic form factor \\
and inelastic neutron scattering}
\author{A. Severing}
  \affiliation{Institute of Physics II, University of Cologne,
   Z{\"u}lpicher Stra{\ss}e 77, D-50937 Cologne, Germany}
\author{F. Givord}
  \affiliation{CEA-Grenoble, DSM/DRFMC/SPSMS/MDN, 38054 Grenoble Cedex 9, France}
\author{J.-X. Boucherle}
  \affiliation{CEA-Grenoble, DSM/DRFMC/SPSMS/MDN, 38054 Grenoble Cedex 9, France}
\author{T. Willers}
  \affiliation{Institute of Physics II, University of Cologne,
   Z{\"u}lpicher Stra{\ss}e 77, D-50937 Cologne, Germany}  
\author{M. Rotter}
  \affiliation{Max Planck Institute CPfS, N{\"o}thnizer Stra{\ss}e 40, 01187 Dresden, Germany}  
\author{Z. Fisk}
  \affiliation{University of California, Irvine, CA, USA}
\author{A. Bianchi}
  \affiliation{D$\acute{e}$partement de Physique and Regroupement Qu$\acute{e}$b$\acute{e}$cois sur les Mat$\acute{e}$riaux de Pointe, Universit$\acute{e}$ de Montr$\acute{e}$al, Montr$\acute{e}$al, Quebec H3C 3J7, Canada}
\author{M.T. Fernandez-Diaz}
  \affiliation{Institut Laue Langevin, 6 rue Horowitz, 38042 Grenoble, France}
\author{A. Stunault}
  \affiliation{Institut Laue Langevin, 6 rue Horowitz, 38042 Grenoble, France}
\author{B.D. Rainford}
  \affiliation{School of Physics and Astronomy, University of Southampton, Southampton, SO17 1BJ, United Kingdom}
\author{J. Taylor}
  \affiliation{ISIS Facility, Rutherford Appleton Laboratory, Chilton, Didcot, Oxfordshire OV11 0QX, United Kingdom}
\author{E. Goremychkin}
  \affiliation{ISIS Facility, Rutherford Appleton Laboratory, Chilton, Didcot, Oxfordshire OV11 0QX, United Kingdom}
\date{\today}

\begin{abstract}
The magnetic form factor of YbInNi$_4$ has been determined via the flipping ratios \textsl{R} with polarized neutron diffraction and the scattering function S(Q,$\omega$) was measured in an inelastic neutron scattering experiment. Both experiments were performed with the aim to determine the crystal-field scheme. The magnetic form factor  clearly excludes the possibility of a $\Gamma_7$ doublet as the ground state. The inelastic neutron data exhibit two, almost equally strong peaks at 3.2 meV and 4.4 meV  which points, in agreement with earlier neutron data, towards a $\Gamma_8$ quartet ground state. Further possibilities like a quasi-quartet ground state are discussed.
\end{abstract}

\pacs{71.27.+a, 75.10.Dg, 78.70.Nx, 75.30.Mb}

\maketitle
\section{Introduction}

YbInNi$_4$ came into the focus of interest when the first order valence transition from a trivalent high temperature to an intermediate low temperature phase was discovered in YbInCu$_4$.\cite{Felner1987} In many rare earth compounds the 4\textsl{f} electrons couple with the conduction electrons, leading to the wealth of properties from magnetic order, to Kondo and Heavy Fermion states with non-conventional superconductivity and/or non-Fermi liquid behaviour, and intermediate valency. Hence the knowledge of the low energy excitations, especially the ground state wave function of the \textsl{4f} electrons involved, is crucial for any further understanding of what mechanism leads to what property. The crystal-field splitting of YbInCu$_4$ is difficult to study directly because of its intermediate valent state at low temperatures. Instead YbInNi$_4$ has been investigated, because it is trivalent over the entire temperature range. However, there are many discrepancies between the different crystal-field proposals \cite{Sarrao1998, Severing1900, Pagliuso2001, Aviani2009} so that we re-investigated the crystal-field scheme of YbInNi$_4$ with polarized neutron diffraction and inelastic neutron scattering. 

At high temperatures YbInNi$_4$ shows Curie-Weiss behaviour with the full Yb$^{3+}$ magnetic moment and orders magnetically at 3 K.\cite{Sarrao1998, selbst} YbInNi$_4$ forms in the cubic C15b Laves structure with T$_d$ symmetry at the Yb site, so that the 8-fold degenerate Hund's rule ground state of Yb$^{3+}$ with J=7/2 and J$_z$=$|\pm7/2\rangle$, $|\pm5/2\rangle$, $|\pm3/2\rangle$, $|\pm1/2\rangle$ is lifted by the crystal-field into two Kramer's doublets $|\Gamma_6$$\rangle$ and $|\Gamma_7$$\rangle$, and one quartet $|\Gamma_8$$\rangle$. In cubic site symmetry the $J_z$ admixture of these crystal-field wave functions is fixed so that the crystal-field scheme is fully determined with the crystal-field transition energies and the sequence of states. In Stevens approximation the transition energies and sequence of states is given by the Stevens parameters. For Yb$^{3+}$ with cubic T$_d$ symmetry the two Stevens parameters $B^{0}_{4}$ and $B^{0}_{6}$ fully describe the crystal-field scheme.\cite{Stevens} 

There has been a controversy about the crystal-field schemes for several years and it became of interest again because of the more recent discovery that applying pressure (2.45 GPa) to YbInCu$_4$ suppresses the valence transition and leads to magnetic order at 2.4~K.\cite{Mito2003, Mushnikov2003, Mito2007} There are several crystal-field propositions around which are briefly summarized in the following: 1) Inelastic neutron scattering data by Severing \textsl{et al.} \cite{Severing1900} showed a double peak structure at low temperatures and the excitations at about 3 and 4~meV were interpreted as transitions from a quartet ground state because the transition matrix element between the two doublets is zero.\cite{Birgeneau} 2) Sarrao \textsl{et al.} concluded from their entropy findings in the specific heat that the ground state should be a doublet and they specified from magnetization data that it should be the $|\Gamma_7\rangle$ state. Best fits were obtained when assuming the second doublet at 5.5~meV, and the $|\Gamma_8\rangle$ quartet at 10.5~meV.\cite{Sarrao1998} 3) From rare earth doped LuInNi$_4$ data of the magnetic susceptibility and from ESR measurements Pagliuso \textsl{et al.} suggest also a $|\Gamma_7\rangle$ ground state with the $|\Gamma_8\rangle$ quartet at 4~meV and the $|\Gamma_6\rangle$ doublet at 9~meV \cite{Pagliuso2001}. 4) Aviani \textsl{et al.} and Park \textsl{et al.} investigated the crystal-field ground state of YbInCu$_4$, where Aviani \textsl{et al}. suppressed the valence transition with 50\% Y doping and Park \textsl{et al.} by applying pressure. Aviani \textsl{et al} describe their specific heat data well with a quasi-quartet ground state which is the inversed scheme of Severing \textit{et al.}, while Park \textsl{et al.} describe their specific heat data under pressure best with the $|\Gamma_8\rangle$ quartet scenario and a total splitting of 28~K when taking the Kondo effect into account\cite{Aviani2009, Park2006} which is in agreement with the neutron proposal for YbInNi$_4$. 

\section{Experimental and Analysis}
These different proposition challenged us to make another attempt to determine the crystal-field scheme of YbInNi$_4$ with neutron scattering. We have measured the low temperature magnetic form factor, which probes directly the ground state wave function as Fourier transform of the spatial distribution of the \textsl{4f} electron moment, i.e. it is directly sensitive to the anisotropy of the crystal-field ground state. Moreover, we have performed inelastic neutrons scattering experiments on a powder sample with emphasis on measuring excitations at low temperatures with a resolution better than in reference~\onlinecite{Severing1900}. Crystals were grown by flux growth and the C15b structure has been verified by powder x-ray diffraction. 

\subsection{Magnetic Form Factor}

The magnetic scattering intensity in an elastic neutron scattering experiment is determined by the magnetic structure factor $F_M$ where 

\begin{equation} 
F_M=\frac{r_0}{2\mu_B}\sum_n|\textbf{m}_n|(\textbf{Q})exp(i\textbf{QR}_n)exp(-W_n)
\end{equation}

and where $\textbf{m}(\textbf{Q})$ is the Fourier transform of the magnetization density at the scattering vector $\textbf{Q}$. The summation is over the \textsl{n} atoms of the magnetic unit cell, $exp(-W_n)$ is the Debye-Waller factor and $R_n$ the position of the $n$th atom, $\mu_B$ is the Bohr magneton, and  $r_0=\gamma\frac{e^2}{mc^2}$ with $\gamma$=-1.92 the gyromagnetic ratio of the neutron. In dipole approximation $\textbf{m}(\textbf{Q})$ depends only on $|\textbf{Q}|$ and can be written as the product of the magnetic moment $\textbf{m}$ and the spherical magnetic form factor $f(|\textbf{Q}|)$, $\textbf{m}(|\textbf{Q}|)$ = $\textbf{m}$$f(|\textbf{Q}|)$. $f(|\textbf{Q}|)$ is listed in text books.\cite{Brown} However, the dipole approximation is not valid for large momentum transfers and/or when the spatial distribution of magnetic moments is strongly anisotropic as for $4f$ moments in the presence of a crystal-field. Then the magnetic scattering intensity is proportional to $|\textbf{m}_\bot(\textbf{Q})|^2$, which is the square of the projection of $\textbf{m}(\textbf{Q})$ perpendicular to $\textbf{Q}$. This introduces in addition to the modulus a \textsl{vector} \textbf{Q} dependence to the magnetic intensities which does not exist in dipole approximation. This vector \textbf{Q} dependence of $\textbf{m}_\bot(\textbf{Q})$ can be used to determine crystal-field ground state wave functions (see e.g. reference \onlinecite{Boucherle2001, Boucherle1985, Rotter}). Here we determined the vector \textbf{Q} dependence of $\textbf{m}(\textbf{Q})$ of Yb$^{3+}$ with the aim to obtain the ground state wave function of YbInNi$_4$. A polarized neutron diffraction experiment was performed on a YbInNi$_4$ single crystal using the D3 diffractometer at the high flux reactor of the Institut Laue-Langevin in Grenoble. The peak intensity of the Bragg reflections which is measured for neutrons polarized parallel and antiparallel to the applied magnetic field, leads to the so-called flipping ratio $R = I^+/I^-$. To deduce the magnetic contribution from $R$, a good knowledge of the nuclear structure factors $F_{\rm N}$ is necessary. This was achieved by an experiment on the same crystal on the 4-circle diffractometer D9 at ILL.
\begin{figure}[]
    \centering
    \includegraphics[width=1\columnwidth]{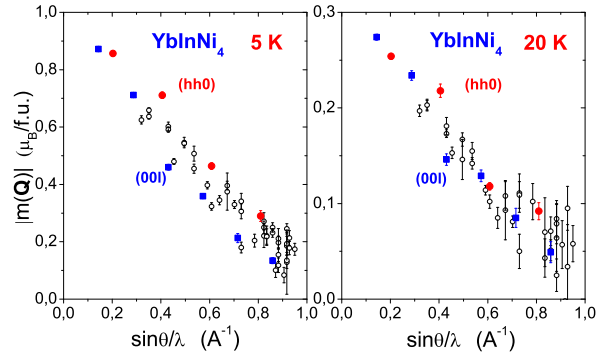}
   \vspace{-8mm}
    \caption{(color online) Ytterbium magnetic amplitudes $|\textbf{m}(\bf{Q})|$ 
    measured with an applied field of 2 T at $T$ = 5~K (a) and $T$ = 20~K (b). The full blue squares correspond to (00l), the full red circles to (hh0), and the black open squares to other reflections with even indices.}
\end{figure}

The nuclear intensity measurements with unpolarized neutrons (D9) were performed at $T$ = 5~K with two different wavelengths $\lambda$ = 0.84~\AA (1146 Bragg reflections) and $\lambda$ = 0.51~\AA (698 Bragg reflections). The crystal structure of our sample was refined to the C15b structure with the program MXD.\cite{MXD} The scattering lengths were taken from the BNL tables \cite{BNL} as 12.43~fm for ytterbium and 10.3~fm for nickel. As the scattering length of indium depends strongly on the wavelength, its real and imaginary parts ($b$ and $b"$, respectively) were calculated for each wavelength from the values given at 1.8~\AA. The total absorption coefficient $\mu$ was also calculated and all these wavelength dependent values are gathered in table I. The nickel atomic position $x_{\rm Ni}$, the Debye-Waller thermal parameters $W_n$, and the extinction parameters, block size $t$ and mosaicity $g$,\cite{ext} were refined, leading to consistent values at both wavelengths. These values are given in table II.

The polarized neutron diffraction experiments (D3) were performed at two temperatures, 5 and 20~K, in a field of H~=~2~T applied along the $[1\bar{1}0]$ axis of the crystal. The selected wavelength was $\lambda$ =0.825~\AA, and two erbium filters were used to suppress the $\lambda/2$ contamination. To optimize the extinction corrections,\cite{extcor1,extcor2} flipping ratios of some particular reflections were also measured at two lower wavelengths $\lambda$ = 0.74~\AA~and $\lambda$ = 0.52~\AA, thus taking advantage of the hot source available on D3.

In YbInNi$_4$, the nuclear and magnetic structure factors are complex. Their imaginary part has two origins: the imaginary scattering length of indium and  the non-centrosymmetric structure of YbInNi$_4$. This second  feature  affects the ytterbium contribution to the structure factor only for Bragg reflections with $h,k,l$ odd since the Yb occupy the \textsl{4c} lattice sites at $\frac{1}{4}\frac{1}{4}\frac{1}{4}$. Note that the expectation value $\textbf{m}_{\bot}(\textbf{Q})$ is real for any \textsl{4f} wave function, so that the magnetic structure factor $F_M$ is real for $h,k,l$ even. For $\textbf{m}(\textbf{Q})$~$\|$~\textbf{H} the flipping ratio $R$ can then be written (see e.g. Ref. \onlinecite{Schweizer}):

\begin{equation} \label{eq1}
R = \frac{F'^2_{\rm N}+F''^2_{\rm N}+2\sin^2{\alpha}F'_{\rm N}F_{\rm M}+\sin^2{\alpha}F^2_{\rm M}}
{F'^2_{\rm N}+F''^2_{\rm N}-2\sin^2{\alpha}F'_{\rm N}F_{\rm M} +\sin^2{\alpha}F^2_{\rm M}}
\end{equation}

$F'_{\rm N}$ and $F''_{\rm N}$ are the real and imaginary parts of the nuclear structure factor $F_{\rm N}$ and   $\alpha$ is the angle between the magnetic moment \textbf{m}(\textbf{Q}) and the scattering vector \textbf{Q}. This formula undergoes appropriate corrections in order to take into account instrumental imperfections (polarization of the incident beam, flipping efficiency) and extinction effects. 

55 non-equivalent reflections were measured at $T$ = 5~K and 49 ones at $T$ = 20~K, up to $\sin\theta/\lambda=0.94~$\AA$^{-1}$. Among them, 13 independent reflections measured at several different wavelength´s were used to refine the extinction parameters. With fixing the Ni positions $x_{\rm Ni}$, the Debye Waller parameters $W_n$, and block size $t$ to the values deduced from the refinements on D9, the mosaicity $g$ was refined to a value in complete agreement with the previous ones (see table II). The final values of the different parameters used for the data analysis are also given in that table II (see reference \onlinecite{extcor1,extcor2}).
\begin{table}[]
\begin{center}
\caption{Values of the indium scattering length (real part $b_
{\rm In}$ and imaginary part $b"_{\rm In}$) and of the total 
absorption coefficient $\mu_{\rm YbInNi_4}$ as a function of 
the wavelength $\lambda$.}

\begin{tabular} {l@{\hskip0.8cm}c@{\hskip0.7cm}c
@{\hskip0.7cm}c@{\hskip0.7cm}c
@{\hskip0.7cm}c}\\ \hline\hline

\hline
Device& $\lambda$&$b_{\rm In}$&$b"_{\rm In}$&$\mu_{\rm YbInNi_4}$ \\ 
&(\AA)&(fm)&(fm)&(cm$^{-1}$)\\ 
\hline
D9&0.84&0.3924&-0.0060&1.42\\ 
&0.51&0.3537&-0.0083&1.21\\ 
\hline
D3&0.825&0.3917&-0.0061&1.42\\ 
&0.74&0.3867&-0.0063&1.34\\ 
&0.52&0.3564&-0.0081&1.21\\ 
\hline
\end{tabular}
\end{center}
\end{table}

\begin{table*}[]
\begin{center}   
\caption{Results of the refinements of the nuclear structure. Values 
of the nickel atomic position $x_{\rm Ni}$, the Debye-Waller thermal 
parameters $W_n$, extinction parameters $t$ and $g$, and $\chi^2$=
$\sum_ip_i(A_i^{obs}-A_i^{calc})^2/N_{obs}-N_{var})$ with $p_i=1/\sigma^2$
and $A_i=I_i$ (intensity) for D9 or $A_i=R_i$ (flipping ratio) for D3.}
\begin{tabular} {l@{\hskip0.8cm}c@{\hskip0.8cm}c@{\hskip0.8cm}
c@{\hskip0.8cm}c@{\hskip0.8cm}c@{\hskip0.8cm}c@{\hskip0.8cm}c@
{\hskip0.8cm}c@{\hskip0.8cm}c} \\ \hline
Device& $\lambda$&$x_{\rm Ni}$&$W_{\rm Yb}$&$W_{\rm In}$&$W_
{\rm Ni}$&$t$&$g$& $\chi^{2}$ \\ 
&(\AA)&&(\AA$^2$)&(\AA$^2$)&(\AA$^2$)&($\mu$m)&(10$^{-4}$rad$^{-1}$)&\\ 
\hline
D9-refined&0.84&0.62573&0.083&0.178&0.140&4.8&0.074&\\ 
&&($\pm$0.00002)&($\pm$0.006)&($\pm$0.011)&($\pm$0.005)&($\pm$0.7)&($\pm$0.006)\\ 
&0.51&0.62583&0.146&0.113&0.179&5.2&0.094&4.5\\ 
&&($\pm$0.00004)&($\pm$0.016)&($\pm$0.033)&($\pm$0.014)&($\pm$2.6)&($\pm$0.023)\\ 
\hline
D3-refined&3$\lambda$&0.62575&0.10&0.15&0.15&5.0&0.094&2.4\\  
&&&&&&&($\pm$0.013)\\ 
\hline
D3-final&3$\lambda$&0.62575&0.10&0.15&0.15&5.0&0.09\\
&&($\pm$0.00005)&($\pm$0.03)&($\pm$0.03)&($\pm$0.03)&&($\pm$0.02)\\ 
\hline
\end{tabular}
\end{center}
\end{table*}
The $F_{\rm M}$ values deduced from the measured flipping ratios are directly related to the Fourier transform of the magnetization density $\textbf{m}(\textbf{Q})$ for the scattering vector $\bf{Q}$ (see eq.(1)). Fig. 1 shows the obtained $|\textbf{m}(\textbf{Q})|$ values at 5~K and 20~K versus $\sin{\theta}/\lambda$ with $|\bf{Q}|$=$4\pi\sin{\theta}/\lambda$. At 5~K there is an anisotropy, i.e. all Bragg reflections lie between the ($hh0$) and the ($00l$)-type ones. The ($00l$) reflection decrease steeper with $\sin{\theta}/\lambda$ than the ($hh0$) ones. At 20~K, this anisotropy has nearly vanished.  

Only these special reflections are drawn in Fig. 2. We have now simulated $|\textbf{m}(\textbf{Q})|$ for the different crystal-field scenarios as suggested by.\cite{Sarrao1998,Severing1900,Pagliuso2001,Aviani2009} The simulation was performed with the program package McPhase \cite{Rotter_1} which includes form factor calculations beyond dipole approximation. Fig. 2 displays the measured $|\textbf{m}(\textbf{Q})|$ values for the ($00l$) and ($hh0$) reflections as full symbols. The value for $\sin{\theta}/\lambda$ = 0 is taken from magnetization. The solid lines are simulations. For the current plot the simulations have been scaled to the value at $\sin{\theta}/\lambda$ = 0. The scaling factors are 1.07, 1.05, and 0.95 for the $|\Gamma_6\rangle$, $|\Gamma_7\rangle$, and $|\Gamma_8\rangle$ simulation. The simulations show that a $|\Gamma_7\rangle$ ground state would yield the wrong anisotropy: For a $|\Gamma_7\rangle$ the ($hh0$) reflections decrease steeper with $\sin{\theta}/\lambda$ than the ($00l$) which is in contradiction to our 
observation. Hence our form factor measurements clearly rule out the possibility of $|\Gamma_7\rangle$ as ground state. In contrast, the simulations for the $|\Gamma_6\rangle$ ground state of the quasi-quartet scenario and the for the $|\Gamma_8\rangle$ quartet both give the correct anisotropy.
\begin{figure}[]
    \centering
    \includegraphics[width=1.0\columnwidth]{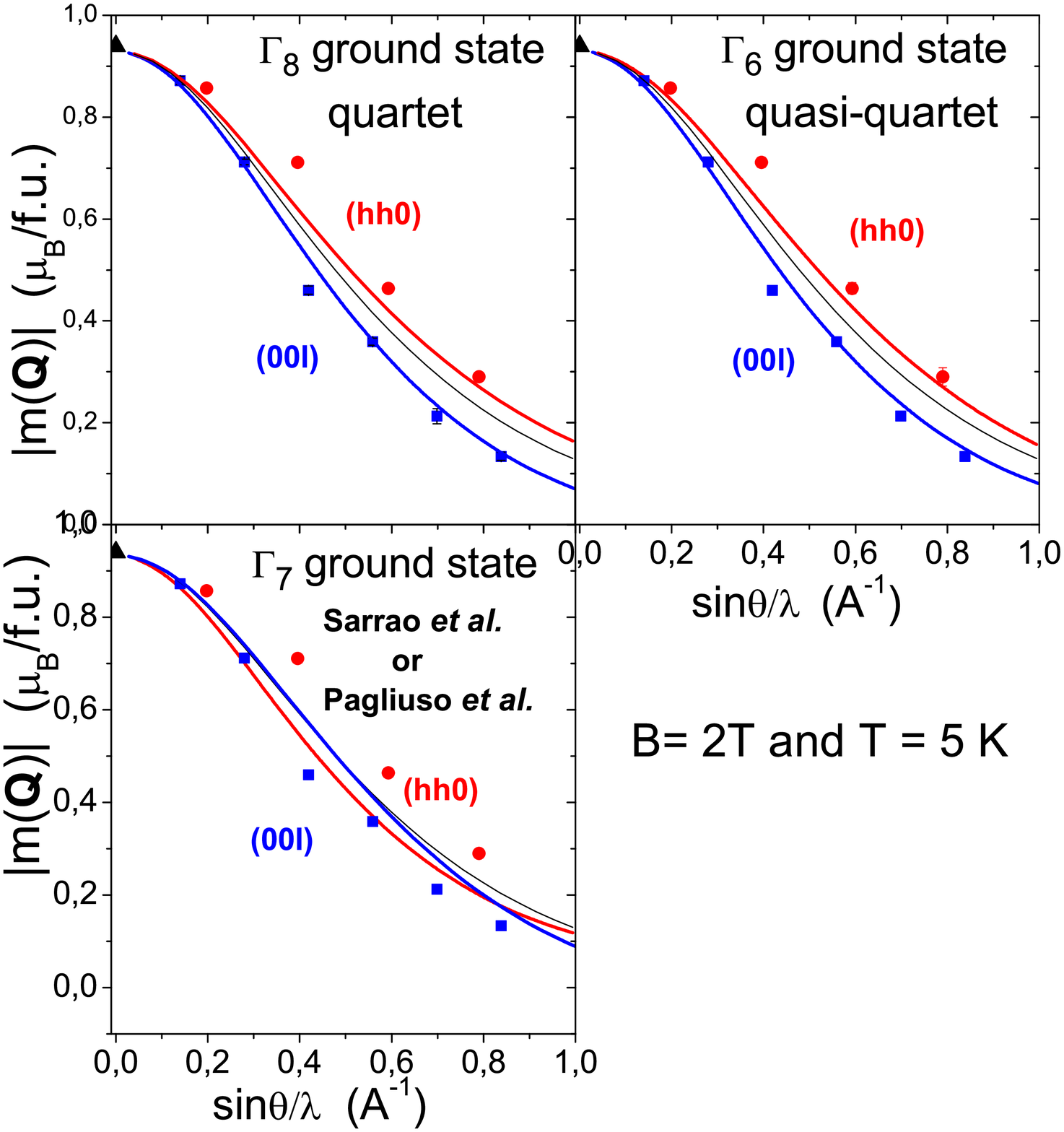}
   \vspace{-8mm}
    \caption{(color online) Comparison of simulated ytterbium magnetic 
    amplitudes $|\textbf{m}(\bf{Q})|$ at $T$ = 5~K for different 
    ground states. The full symbols are the measured values for
    ($00l$) (blue squares) and ($hh0$) (red circles), the lines are simulations for the 
    ($00l$) (blue) and ($hh0$) (red) directions for the different crystal-field propositions. The black line 
    corresponds to the spherical form factor. The values for $\sin{\theta}/\lambda$ = 0 (black triangles) come   from magnetization.}
\end{figure}
\subsection{Inelastic Neutron Scattering}
Inelastic neutron scattering experiments on polycrystalline samples are the most common technique to determine crystal-field excitations in rare earth compounds. Here we present inelastic neutron data of polycrystalline YbInNi$_4$ which were taken at the time-of-flight spectrometer MARI at the pulsed neutron source ISIS with an incident energy of 12~meV and a resolution of 0.4~meV (FWHM) at elastic scattering. Detectors from 2$\theta$ = 8$^{\circ}$ to 52$^{\circ}$ are grouped together, resulting in an averaged Q vector at elastic scattering of 1.34~{\AA}. The sample was mounted in a cryostat and data were taken in the magnetically ordered phase at 2~K and in the paramagnetic phase at 5~K, 20~K, and 40~K. 

\begin{figure}[]
    \centering
    \includegraphics[width=1.0\columnwidth]{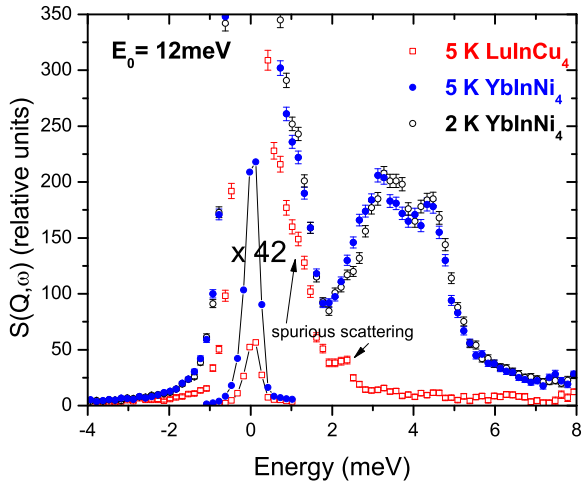}
   \vspace{-8mm}
    \caption{(color online) Scattering function of YbInNi$_4$ for $|\textbf{Q}|$ in between 0.35~{\AA}$^{-1}$ and 2.3~{\AA}$^{-1}$ at 2~K (open black circles) and 5~K (full blue circles), and of LuAuCu$_4$ (open red squares), also at 5~K. The insets represent the elastic scattering divided by 42. The extra scattering in the YbInN$_4$ data between 2.5~meV and 6~meV is attributed to magnetic scattering.}
\end{figure}

\begin{figure}[]
    \centering
    \includegraphics[width=1.0\columnwidth]{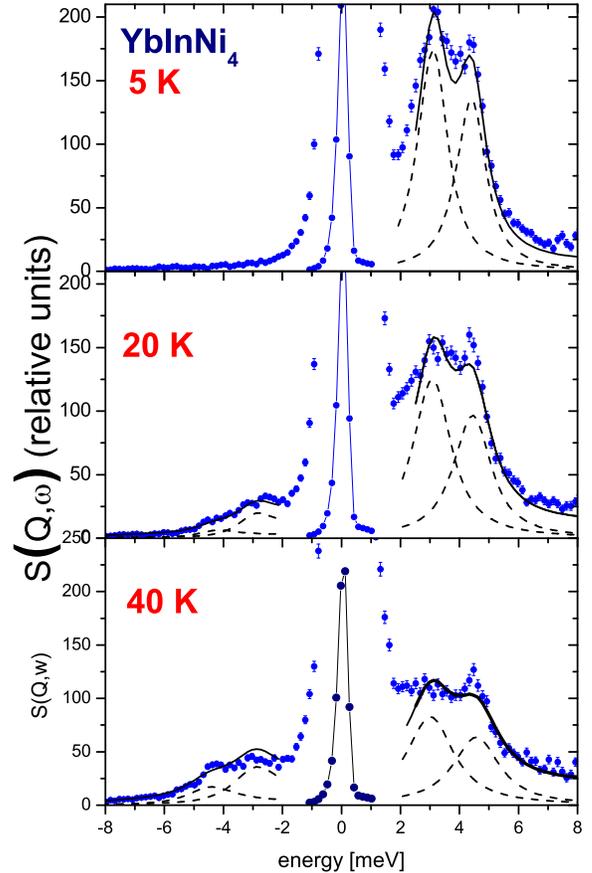}
   \vspace{-8mm}
    \caption{(color online) Temperature evolution of the scattering function of YbInNi$_4$ for 12~meV incident energy and $|\textbf{Q}|$ in between 0.35~{\AA}$^{-1}$ and 2.3~{\AA}$^{-1}$. The insets represent the elastic scattering divided by 42, the dashed lines crystal-field excitations resulting from a crystal-field fit, and the solid lines are the total fitted magnetic scattering. Note, the fitting region was restricted to energy transfers below -2~meV and above 2.5~meV to avoid the strong elastic and the spurious scattering at the tail of the elastic line.}
\end{figure}

Fig. 3 shows the scattering function $S(Q,\omega)$ versus energy transfer for YbInNi$_4$ at 2 and 5~K, and of the non-magnetic isostructural compound LuAuCu$_4$, also at 5~K. The data are corrected for detector efficiency, absorption, and are scaled to sample mass so that the scattering  intensities are comparable. Due to the smaller nuclear cross-section of LuAuCu$_4$ with respect to YbInNi$_4$ the elastic scattering of the latter is stronger. The Lu data prove that phonon scattering is negligible in the present energy window and detector grouping. The extra scattering of the YbInNi$_4$ sample between 2.5~meV and 6~meV is therefore identified as magnetic scattering. The shoulder which appears on the neutron energy loss side of the elastic line appears in the spectra of the magnetic and non-magnetic sample, and is attributed to spurious Bragg scattering in either sample which has been reflected from the cryostat walls, reaching the detectors time delayed, and therefore in an inelastic channel. Since it is difficult to account for this quantitatively and since we know from previous high resolution data \cite{Severing1900} that the quasielastic scattering of YbInNi$_4$ is too narrow to be resolved in the present experiment we concentrate on the inelastic scattering above 2.5~meV and on the neutron energy 
gain side for T = 20~K and 40~K. 

The magnetic scattering intensity between 2.5~meV and 6~meV consists clearly of two lines. At 5 K these lines can be fitted with two Lorentzians centered at about 3.2$\pm$0.1~meV and 4.4$\pm$0.1~meV. In the magnetically ordered phase at 2 K the lower crystal-field excitation appears at slightly larger energy transfers with respect to 5~K which is  probably due to the influence of the magnetic order at 2~K. The line widths remain unchanged.

Fig. 4 shows the temperature evolution of the scattering function in the paramagnetic phase from 5~K, 20~K to 40~K. The double peak structure survives up to 40~K without a shift in energy, although the excitations become broader as temperature rises. The present incoming neutron energy of 12~meV provides an energy window up to 10~meV. At none of the temperatures magnetic scattering was detected at energy transfers larger than than 4.4~meV.

The scattering function S(Q,$\omega$) can exhibit two magnetic excitations at 2 and 5~K only if\\
a) the ground state is the $|\Gamma_8\rangle$ quartet and two ground state excitations 
   into the $|\Gamma_6\rangle$ and $|\Gamma_7\rangle$ doublets take place, or\\
b) the ground state is a quasi-quartet and the two doublets which are 1 meV apart are sufficiently populated, or\\  
c) if the crystal-field states are Zeeman split due to the proximity of the magnetically ordered state. Then even a doublet could give rise to two ground state excitations.\\
d) Or if some structural distortion is present, so that the site symmetry is no longer cubic.\\
\begin{figure}[]
    \centering
    \includegraphics[width=1.0\columnwidth]{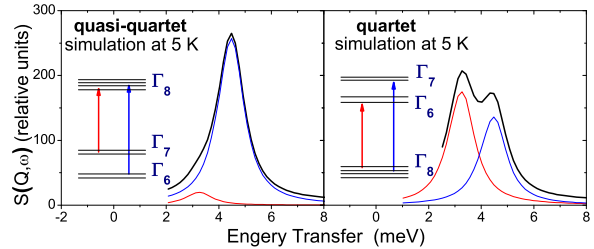}
   \vspace{-8mm}
    \caption{(color online) Simulated inelastic scattering function at 5 K for a quasi-quartet (left) and quartet (right) ground state crystal-field model.}
\end{figure}  
Proposition a) is based on the fact that the transition matrix element between the two doublets is zero.\cite{Birgeneau} This new set of data is also, like the previously reported neutron data, well described with a $|\Gamma_8\rangle$ quartet ground state. For a quantitative analysis the magnetic form factor has to be taken into account. Here, for the purpose of describing non dispersive excitations in a powder sample it is sufficient use the spherical, text book magnetic form factor $f(|\textbf{Q}|)$.\cite{Brown} The result of a quantitative crystal-field analysis where all three temperatures were fitted simultaneously is shown in Fig. 4 (see black lines). For both excitations the same line width was assumed. The fitted line widths increase with temperature from 0.54$\pm$0.05~meV at 5~K to 0.7$\pm$0.05~meV at 20~K and 0.9$\pm$0.08~meV at 40~K. The best fit was obtained for 
 $B^{0}_{4} = -1.1 10^{-3}~meV$ and   $B^{0}_{6} = -8.88 10^{-5}~meV$,
which is consistent with reference \onlinecite{Severing1900} but seemingly in conflict with the entropy findings of the specific heat\cite{Sarrao1998}. Although b), the quasi-quartet scenario $\grave{a}$ la Aviani \textsl{et al.}\cite{Aviani2009} is tempting, but a simple simulation of the scattering function for a quasi-quartet ground state excludes this possibility as an explanation for the inelastic neutron data of YbInNi$_4$: neither at 2 nor at 5~K the first excited doublet at 1~meV is sufficiently populated to give rise to two almost equally strong excitations. This is demonstrated in Fig. 5 where the result of a crystal-field simulation with a quasi-quartet ground state is shown for 5~K. Possibility c) of a molecular field split doublet ground state could give rise to two excitations at low temperatures, but for e.g. a $|\Gamma_6\rangle$ ground state the same molecular field of $\approx$0.33~meV would have to be assumed for the magnetically ordered and the paramagnetic phase. Another difficulty with this assumption is that at higher temperatures, when no Zeeman splitting and intermixing of states due to a molecular field is present, the spectra should be shifted in energy with respect to the low temperature data. This has not been observed. Possibility d) of a structural distortion can be ruled out from our structural analysis which is part of the form factor measurement and from X-ray data which were taken in order to verify sample quality. Hence, this new set of neutron data confirms the previous inleastic neutron scattering  results of a $|\Gamma_8\rangle$ quartet ground state.

\section{Discussion}

The observed anisotropy of the magnetic form factor is clearly not compatible with the spatial distribution of a $|\Gamma_7\rangle$ ground state although magnetization and susceptibility measurements \cite{Sarrao1998, Pagliuso2001} favour $|\Gamma_7\rangle$. This demonstrates the difficulty to determine crystal-field ground states from magnetic measurements where thermal averaging, exchange, possible Kondo interactions, etc. enter the description but are not easily taken care of. The magnetic form factor, however, cannot distinguish between $|\Gamma_6\rangle$ and $|\Gamma_8\rangle$ because the projection of the orbitals of these states on to the scattering plane is very much the same. A $|\Gamma_6\rangle$ ground state would be in agreement with the entropy findings of the specific heat \cite{Sarrao1998} but as has been pointed out in the upper paragraph, it fails to explain the double peak structure in the inelastic neutron scattering data. All inelastic neutron scattering data exhibit a double peak structure at all temperatures and can be analyzed consistently with a $|\Gamma_8\rangle$ quartet ground state and two ground state excitations. From the present experiment with 12 meV incident energy and from the previous one with 17 meV we can conclude further that there is no additional ground state excitation within an energy window up to 15 meV. 

However, it should be mentioned that additional peaks in the inelastic neutron data have been observed for example in CeAl$_2$.\cite{Loewenhaupt} Here strong electron phonon coupling gives rise to an additional peak in S(Q,$\omega$). But this usually happens at higher energy transfers of acoustic zone boundary or optical phonon branches because the observation of such a bound state  requires a certain phonon density of states which is not given in the range where acoustic branches rise steeply. For CeAl$_2$ the bound state is observed at 10 meV which coincides with the first peak in the phonon density of states. Ytterbium is not much heavier than cerium so that it is unlikely to have a flat phonon branch in the energy range of 3-4 meV. Hence we consider it unlikely that electron phonon coupling gives rise to an extra excitation. 

We consider the quartet ground state in YbInNi$_4$ as the most likely solution. This makes us speculate whether YbInNi$_4$ is a candidate for quadrupolar order as e.g. in CeB$_6$. We further believe that the magnetic exchange interaction in YbInNi$_4$ is predominantly antiferromagnetic\cite{selbst} and not ferromagnetic as suggested in reference \onlinecite{Sarrao1998} which then leads us to speculate whether there is magnetic frustration due to the \textsl{fcc} crystal-structure of YbInNi$_4$ where the Yb ions are arranged in corner sharing tetrahedra. We therefore do not think that the entropy findings of \textsl{Rln2} by Sarrao \textsl{et al.} are necessarily in conflict with a quartet ground state. 

\section{Summary}

The crystal-field scheme of YbInNi$_4$ has been determined with magnetic form factor and inelastic neutron scattering measurements. The asymmetry of the magnetic form factor clearly rules out a $|\Gamma_7\rangle$ doublet as ground state (see Fig. 2) and would be in line with either a $|\Gamma_6\rangle$ or $|\Gamma_8\rangle$ ground state. A quasi-quartet ground state, where the two doublets are 1 meV apart, would not yield two almost equally strong excitations in the inelastic neutron data at 5 K since the first excited state would not be sufficiently populated (see Fig. 5). The inelastic neutron data confirm the existence of two crystal-field excitations at 3.2 and 4.4 meV at low temperatures and due to selection rules this is only possible for a $|\Gamma_8\rangle$ quartet ground state. 

\section{Acknowledgment}
We would like to thank Andrew Boothroyd for helpful discussions. T.W. is supported by the Bonn-Cologne Graduate School of Physics and Astronomy. 


\begin{thebibliography}{99}

\bibitem{Felner1987} I. Felner {\it et al.}, Phys. Rev. B {\bf 35}, 6956 (1987). 

\bibitem{Sarrao1998} J.L. Sarrao {\it et al.}, Phys. Rev. B {\bf 57}, 7785 (1998).

\bibitem{Severing1900} A. Severing,E. Gratz, B.D. Rainford, K. Yoshimura, 
Physica B {\bf 163}, 409 (1990).

\bibitem{Pagliuso2001} P.G. Pagliuso {\it et al.}, Phys. Rev. B {\bf 63} 144430 (2001).

\bibitem{Aviani2009} I. Aviani, M. Ocko, D. Starescinic, K. Biljakovic, A. Loidl, J. Hemberger, J.L. Sarrao, Phys. Rev. B \textbf{79}, 165112 (2009).

\bibitem{selbst} T. Willers and A. Severing \textsl{et al.}, to be published.

\bibitem{Stevens} K.W. H. Stevens, Proc. Phys. Soc. A {\bf 65}, 209 (1951).

\bibitem{Mito2003} T. Mito, T. Koyama, M. Shimoide, S. Wada, T. Muramatsu, T.C. Kobayashi, and J.L. Sarrao,
 Phys. Rev. B {\bf 67}, 224409, (2003). 

\bibitem{Mushnikov2003} T. Goto, E.V. Rozenfeld, K. Yoshimura, W. Zhang, M. Yamada, and H. Kageyama,
J. Phys. Condens. Matter \textbf{15}, 2811 (2003)

\bibitem{Mito2007} T. Mito, M. Nakamura, M. Otani, T. Koyama, S. Wada, M. Ishizuka, 
M.K. Forthaus, R. Lengsdorf, M.M. Abd-Elmeguid, and J.L. Sarrao. Phys. Rev. B {\bf 75}, 134401 (2007).

\bibitem{Birgeneau} R.J. Birgeneau, J. Phys. Chem. Sol. \textbf{33}, 59 (1972)

\bibitem{Park2006} T. Park, V. A. Sidorov, J. L. Sarrao, and J. D. Thompson, Phys. Rev. Lett. {\bf 96}, 
046405 (2006).

\bibitem{Brown} P.J. Brown, Magnetic form factors, 
chapter 4.4.5, International tables for crystallography 
vol. C (A. J. C. Wilson, ed.), pp. 391-399.

\bibitem{Boucherle2001} J.X. Boucherle, F. Givord, S. Raymond, J. Schweizer, 
E. Lelievre-Berna and G. Fillion, J. Phys. Cond. Mat. B \textbf{13}, no 48, 10901 (2001)

\bibitem{Boucherle1985} J.X. Boucherle and J. Schweizer, Physica B \textbf{130}, 337 (1985)

\bibitem{Rotter} M. Rotter and A.T. Boothroyd, Phys. Rev. B \textbf{79}, 140405 (2009)

\bibitem{MXD} P. Wolfers, J. Appl. Cryst. \textbf{23} 554 (1990).

\bibitem{BNL} V.F.  Sears, Neutron News \textbf{3} 26 (1992).

\bibitem{ext} P. Becker and P. Coppens, Acta Crystallogr. \textbf{A30} 129 (1974).

\bibitem{extcor1} M. Bonnet, A. Delapalme, P. Becker and H. Fuess, 
Acta Crystallogr. \textbf{A32} 945 (1976).

\bibitem{extcor2} J-X. Boucherle and J. Schweizer, J. Magn. Magn. 
Mater. \textbf{24}, 308 (1981).

\bibitem{Schweizer} J. Schweizer in \textsl{Neutron Scattering form Magnetic Materials}, ed. by T. Chatterji, Elsevier B.V. 2006, p. 153.

\bibitem{Rotter_1} M. Rotter \textsl{et al.}, McPhase, a soft ware package for 
calculation of the phase diagram and magnetic properties of rare earth systems 
(2002-2008), available at http://www.mcphase.de

\bibitem{Loewenhaupt} M. Loewenhaupt and P. Fulde, Advances in Physics, \textbf{34}, No. 5, 589-661 (1986).

\end{thebibliography}
\end{document}